\documentclass[twocolumn,showpacs,preprintnumbers,amsmath,amssymb]{revtex4}
\usepackage{graphicx}% Include figure files
\usepackage{dcolumn}% Align table columns on decimal point
\usepackage{bm}% bold math

\usepackage{amssymb}

\begin{document}

%\preprint{APS/123-QED}

\title{Dirac observables and boundary proposals in quantum cosmology}

\author{S. Jalalzadeh}
\email{s-jalalzadeh@sbu.ac.ir}
 \affiliation{Department of Physics, Shahid Beheshti University, G. C., Evin, 19839 Tehran, Iran.}
\author{P. V. Moniz}
\email{pmoniz@ubi.pt}
 \affiliation{Centro de Matem\'atica e Aplica\c c\~oes - UBI, Covilh\~a, Portugal,\\
 Departamento de F\'isica, Universidade da Beira Interior, 6200 Covilh\~a, Portugal.}

\date{\today}

\begin{abstract}
We study the reduced phase space quantization of a closed Friedmann Universe, where matter content is constituted by two (no-interacting) fluids, namely dust (or cold dark matter) and radiation. It is shown that, for this particular model, specific
 boundary conditions can be related to the algebra of Dirac observables.    
\end{abstract}

%\keywords{Suggested keywords}
\maketitle
\section{Introduction}

The Wheeler-De Witt (WDW) equation is an important element in quantum cosmology,   determining a wave
function for the Universe \cite{A}. It is constructed using the ADM decomposition of the
spacetime manifold in the Hamiltonian formalism of general relativity
\cite{ADM}. However, the WDW quantum geometrodynamics has many technical
and conceptual challenges \cite{K}:  the problem of time \cite{Problem}, the problem of observables,
 factor ordering issues \cite{O}, the global structure of spacetime manifold and the problem of boundary conditions (for more details, see
\cite{A}, \cite{K} and \cite{B}). 

On the other hand, the problem of observables is closely related to the problem
of time \cite{A,Problem}. Let us be more concrete. According to  Dirac \cite{Dirac}, the observables of a theory are those quantities
which have vanishing Poisson brackets at the classical level and satisfy adequate quantum
commutators at the quantum regime,  in the presence of constraints. Regarding
general relativity (GR), it must be pointed that this theory is invariant under the group of diffeomorphism of hyperbolic spacetime
manifold. Therefore, the Hamiltonian formalism of GR contains first class constraints, namely the Hamiltonian
and momentum constraints. This  leads to the conclusion that all
GR Dirac observables should be time independent.

On the other hand, the issue of boundary conditions for the wave function of the Universe has been one
of the most active areas of quantum cosmology \cite{A,K,B}. 
Two leading lines for the WDW quantization  are the  no-boundary proposal \cite{HH} and the tunneling proposal \cite{Vilenkin}. Two other proposals  have been used, defining, through mathematical expressions, explicit procedures
to deal with the presence of classical singularities. More precisely, the wave function should vanish at the classical singularity $\psi(0)=0$ (De Witt boundary condition) \cite{DeWitt}, or its derivative with respect to the scale factor vanishes at
the classical singularity $\psi'(0)=0$ \cite{Sin}. All those 
boundary conditions are {\it ad hoc} chosen,  with some particular physical intuition in mind \cite{Boj}, \cite{A}, \cite{B}, but they are not part of the dynamical law. However, 
according to  De Witt ``the constraints are everything'' \cite{DeWitt} i.e.,
nothing else but the constraints should be needed. 

A pertinent question that may emerge in the context of the previous paragraph
in following: Can a  relation between
the constraints (that are present and whose algebra characterize GR) and
the allowed boundary conditions  be established? If there is such a relation, then
 boundary conditions could be related to the set of possible Dirac observables. Our aim is to show that in the closed homogeneous and isotropic Universe filled with cold dark
matter (dust) and radiation, there is a hidden symmetry, which by means of
a Dirac observable, allows  boundary conditions to be present as  part of a dynamical law. This paper is organized as follows: Our model is presented in section II.
Its quantization and argumentations towards the claim indicated in the abstract,
is provided in section III, which is constituted by three subsections. We are aware that the herein setting is rather restrictive and we will elaborate
more about our choices in section IV.   
\section{The Model}

One of the simplest models in  quantum cosmology is the homogeneous and isotropic FLRW minisuperspace. The line element of FLRW geometry for the closed Universe
is defined by
\begin{eqnarray}\label{1}
ds^2=-N^2(\eta)d\eta^2+a^2(\eta)d\Omega^2_{(3)},
\end{eqnarray}
where $d\Omega^2_{(3)}$ is the standard line element on the unit three-sphere.
The action functional corresponding to the line element (\ref{1}) for a 
gravitational sector, described by GR plus a  matter content (in the form of a perfect fluid with barotropic equation of state $\rho=\gamma
p$) is \cite{Hawking1}
\begin{eqnarray}\label{2}
\begin{array}{cc}
S=\frac{M_{\text{Pl}}^2}{2}\int_{\cal M}\sqrt{-g}Rd^4x+\\
\\
+M_{\text{Pl}}^2\int_{\partial{\cal
M}}\sqrt{g^{(3)}}Kd^3x
-\int_{\cal M}\sqrt{-g}\rho d^4x\\
\\=6\pi^2M_{\text{Pl}}^2\int\left(-\frac{a\dot{a}^2}{N}+Na\right)d\eta-2\pi^2\int
Na^3\rho d\eta,
\end{array}
\end{eqnarray}
 where $M_{\text{Pl}}^2=\frac{1}{8\pi G}$ is the reduced Planck's mass in
 natural units, ${\cal M}=I\times S^3$ is the spacetime manifold, $\partial{\cal
 M}=S^3$, $K$ is the trace of the extrinsic curvature of the spacetime boundary
 and overdot denotes differentiation with respect to $\eta$. To obtain the correct dynamical equations from a variation of an action such as (\ref{2}), it is necessary to
require the current vector of the fluid to be covariantly conserved \cite{Hawking1}. Consequently,  for a  Universe filled by dust (cold dark matter) and radiation,
non-interacting, which we will taken herein as the matter content of our model, we have
 \begin{eqnarray}\label{3}
 \rho:=\rho_{m}+\rho_\gamma=\rho_{0m}\left(\frac{a}{a_0}\right)^{-3}+\rho_{0\gamma}\left(\frac{a}{a_0}\right)^{-4},
 \end{eqnarray}
 where $\rho_m$ and $\rho_\gamma$ denote the energy densities of dust and
 radiation fluids, respectively \footnote{Quantum
cosmologies with a perfect fluid matter content were investigated previously in \cite{Shahram}. In particular, the FLRW quantum cosmology of a dust or radiation dominated Universe was
widely investigated in the literature. For example, the causal interpretation was studied in \cite{Pinto1}
and \cite{Pinto2}, the problem of time was discussed
in \cite{Chin} and avoiding the Big-Bang singularity in \cite{Narlikar}.}. Hence the corresponding Lagrangian will be
 \begin{eqnarray}\label{4}
 {\cal L}=6\pi^2M_{\text{Pl}}^2\left(-\frac{a\dot{a}^2}{N}+Na\right)-MN-{\cal
 N}_\gamma\frac{N}{a},
 \end{eqnarray}
 where 
 \begin{eqnarray}
 \begin{cases}
 M=\int_{\partial {\cal M}}\sqrt{g^{(3)}}\rho_{0m}a_0^3d^3x,\\
 
 {\mathcal N}_\gamma=\int_{\partial {\cal M}}\sqrt{g^{(3)}}\rho_{0\gamma}a_0^4d^3x.
 \end{cases}
 \end{eqnarray}
  $M$ is the total mass of the dust content of the Universe and ${\mathcal N}_\gamma$ could be related
 to the total entropy of radiation: for radiation, the energy density $\rho_\gamma$,
 the number density $n_\gamma$, the entropy density $s_\gamma$ and scale
 factor are related to the temperature via $\rho_\gamma=\frac{\pi^2}{30}T^4$,
 $n_\gamma=\frac{2\zeta(3)}{\pi^2}T^3$, $s_\gamma=\frac{4\rho_\gamma}{3T}$
 ans $a\propto\frac{1}{T}$ \cite{Mukhanov}. Consequently we obtain ${\mathcal
 N}_\gamma=(\frac{5\times3^5}{2^{10}\pi^4})^\frac{1}{3}S_\gamma^\frac{4}{3}$,
 where $S_\gamma$ is the total entropy of radiation.
 If we redefine the lapse function $N$ and scale factor $a$ as
 \begin{eqnarray}\label{6}
 \begin{cases}
 a(\eta)=x(\eta)+\frac{M}{12\pi^2M_{\text{Pl}}^2}:=x-x_{0},\\
 
 N(\eta)=12\pi^2M_{\text{Pl}}a(\eta)\tilde{N},
 \end{cases}
 \end{eqnarray}
 the Lagrangian (\ref{4}) will be
 \begin{eqnarray}\label{7}
 {\mathcal L}=-\frac{1}{2\tilde{N}}M_{\text{Pl}}\dot{x}^2+\frac{\tilde{N}}{2}M_{\text{Pl}}\omega^2x^2-{\mathcal
 E}\tilde{N},
 \end{eqnarray}
 where
 \begin{eqnarray}\label{8}
 \begin{cases}
 {\mathcal E}=\frac{M^2}{2M_{\text{Pl}}}+12\pi^2{\mathcal N}_\gamma M_{\text{Pl}},\\
 
 \omega=12\pi^2M_{\text{Pl}}.
 \end{cases}
 \end{eqnarray}
 To construct the Hamiltonian of the model, note that the momenta conjugate to $x$ and the primary constraint are given by
 \begin{eqnarray}\label{9}
 \begin{cases}
 \Pi_x=\frac{\partial{\mathcal L}}{\partial\dot{x}}=-\frac{\tilde{N}}{M_{\text{Pl}}}\dot{x},\\
 
 \Pi_{\tilde{N}}=\frac{\partial{\mathcal L}}{\partial\dot{\tilde{N}}}=0.
 \end{cases}
 \end{eqnarray}
 Hence, in terms of the conjugate momenta, the Hamiltonian corresponding to
 (\ref{7}) is
 \begin{eqnarray}\label{10}
 {\mathcal H}=-\tilde{N}\left[\frac{1}{2M_{\text{Pl}}}\Pi^2_x+\frac{1}{2}M_{\text{Pl}}\omega^2x^2-{\mathcal
 E}\right].
 \end{eqnarray}
 Because of the existence of constraint (\ref{9}), the Lagrangian of the system is singular and the total Hamiltonian can be constructed by adding to ${\mathcal H}$ the primary constraints multiplied by
arbitrary functions of time, $\lambda$,
\begin{eqnarray}\label{11}
{\mathcal H}_T=-\tilde{N}\left[\frac{1}{2M_{\text{Pl}}}\Pi^2_x+\frac{1}{2}M_{\text{Pl}}\omega^2x^2-{\mathcal
 E}\right]+\lambda\Pi_{\tilde{N}}.
\end{eqnarray} 
 The requirement that the primary constraint should hold during the evolution of the system means that
\begin{eqnarray}\label{12}
\dot\Pi_{\tilde{N}}=\{\Pi_{\tilde{N}},{\mathcal H}_T\}\approx0,
\end{eqnarray}
which leads to the secondary (Hamiltonian) constraint
\begin{eqnarray}\label{13}
H:= \frac{1}{2M_{\text{Pl}}}\Pi^2_x+\frac{1}{2}M_{\text{Pl}}\omega^2x^2-{\mathcal
 E}\approx0.
\end{eqnarray}
In addition, the constraint (\ref{13}) requires a gauge-fixing condition, where a
possibility  is $\tilde{N}=$constant. If we choose the gauge of $\tilde{N}=1/\omega$
and that for  the canonical variables satisfying the Poisson algebra $\{x,\Pi_x\}=1$,
we find the Hamilton equations of motion
\begin{eqnarray}\label{14}
\begin{cases}
\dot{x}=-\frac{1}{\omega M_{\text{Pl}}}\Pi_x,\\

\dot{\Pi}_x=\omega M_{\text{Pl}}x.
\end{cases}
\end{eqnarray}
Using  the Hamiltonian constraint (\ref{13}),
we can easily find the well known solution of a closed Universe
\begin{eqnarray}\label{15}
\begin{cases}
a(\eta)=\frac{a_{\text{Max}}}{1+\sec\phi}\left[1-\sec\phi\cos(\eta+\phi)\right],\\
a_{\text{Max}}:=\frac{M}{12\pi^2M_{\text{Pl}}^2}+\left(\frac{2{\mathcal E}}{M_{\text{Pl}}\omega^2}\right)^{\frac{1}{2}},\\
\cos\phi:=\frac{M}{\sqrt{2{\mathcal E}M_{\text{Pl}}}},
\end{cases}
\end{eqnarray} 
where $a_{\text{Max}}$ is the maximum radius of the Universe and it is assumed that the initial singularity occurs at $\eta=0$. 

%%%%%%%%%%%%%%%%%%%%%%%%%%%%%%%%%%%%%%%%%%%%%%%%%%%%%%%%%%%%%%%%%%%%%%%%%%%%%%%%%%%%%
\section{ Quantization and Dirac Observables}
\subsection{Standard quantization}
The standard quantization
of this simple system is accomplished straightforwardly in the coordinate
representation  $\hat x=x$ and $\hat \Pi_x=-i\partial_x$.
Then the Hamiltonian constraint (\ref{13}) becomes the WDW
equation for the wave function of the Universe
\begin{eqnarray}\label{28}
-\frac{1}{2M_{\text{Pl}}}\frac{d^2\psi}{dx^2}+\frac{1}{2}M_{\text{Pl}}\omega^2x^2\psi(x)={\mathcal
E}\psi(x).
\end{eqnarray}
Note that the classical solution (\ref{15}) has a singularity at $x=x_0$.
In this context, for the WDW quantization of our model,  we will  assume
 wave functions defined on
the $(x_0,\infty)$ domain, such that boundary conditions
will lead to a self-adjoint Hamiltonian. This therefore suggests us to use
those wave functions which satisfy one of the following boundary conditions:
either De Witt boundary condition
\begin{eqnarray}\label{29}
\psi(x)|_{x=x_0}=0,
\end{eqnarray}
to avoid the singularity at $x=x_0$, or
\begin{eqnarray}\label{29a}
\left(\frac{d\psi}{dx}+\alpha\psi\right)|_{x=x_0}=0,
\end{eqnarray}
where $\alpha$ is a arbitrary constant. As pointed out by Tipler \cite{Tipler}, were condition
(\ref{29a}) chosen then the constant $\alpha$ would be a new fundamental constant of theory. To avoid this new fundamental
constant, we set it to be zero
\begin{eqnarray}\label{29b}
\frac{d\psi}{dx}|_{x=x_0}=0.
\end{eqnarray} 
Using boundary conditions (\ref{29}) or (\ref{29b}),
we obtain normalized oscillator states with eigenvalues ${\mathcal
E}_n=\omega(n+1/2)$, where $n$ is an even or odd  integer, corresponding to
the above boundary conditions (\ref{29}) and (\ref{29b}), respectively.  Hence, using definition (\ref{8}), we obtain
\begin{eqnarray}\label{30}
\begin{cases}
\left(\frac{M}{M_{\text{Pl}}}\right)^2+24\pi^2{\mathcal N}_\gamma=24\pi^2(n+\frac{1}{2}),\\
\psi_n=\left(\frac{\sqrt{M_{\text{Pl}}\omega}}{\sqrt{\pi}2^nn!}\right)^{\frac{1}{2}}H_n(\sqrt{M_{\text{Pl}}\omega}a)\exp{(-\frac{1}{2}M_{\text{Pl}}\omega
a^2)}.
\end{cases}
\end{eqnarray}
As we know, the existence of normalized eigenfunction is directly related
to the existence of the maximum classical radius of a closed Universe \cite{DeWitt}.
Moreover, expression (\ref{30}) suggests  that the mass of dust (dark matter)
and the entropy of radiation are intertwined through a quantization rule.
\subsection{Reduced phase space and observables}
As is well known, GR is invariant under the group of diffeomorphisms of the spacetime manifold $\mathcal{M}$. The main consequences of such a diffeomorphism invariance are that the Hamiltonian can be expressed as a sum of constraints and that any observable must commute with these constraints. An observable is a function on the constraint surface, such that is invariant under the gauge transformations generated by all of the first class constraints. By
a first class constraint we mean a phase space function with the property that it has weakly vanishing Poisson bracket with all constraints. As an example, the momentum and Hamiltonian constraints are always first class, see (\ref{9}) and (\ref{13}). The Hamiltonian and momentum constraints in GR are generators of the corresponding gauge transformations,
and so, a function on the phase space is an observable if it has weakly vanishing Poisson brackets with the first class constraints. To find gauge invariant observables, we can proceed as follows. The unconstrained phase space $\Gamma$
of the model is $\mathbb{R}^2$, with global canonical coordinates $(x,\Pi_x)$
with Poisson structure $\{x,\Pi_x\}=1$. Let us define on $\Gamma$ the complex-valued
functions
\begin{eqnarray}\label{20}
\begin{cases}
C:=\sqrt{\frac{M_{\text{Pl}}\omega}{2}}\left(x+i\frac{\Pi_x}{M_{\text{Pl}}\omega}\right),\\
C^*:=\sqrt{\frac{M_{\text{Pl}}\omega}{2}}\left(x-i\frac{\Pi_x}{M_{\text{Pl}}\omega}\right).
\end{cases}
\end{eqnarray}
The set $S=\{C,C^*,1\}$ is closed under the Poisson bracket, $\{C,C^*\}=-i$
and every sufficiently differentiable function on $\Gamma$ can be expressed
in terms of $S$. Therefore, the Hamiltonian can be viewed  as
\begin{eqnarray}\label{21}
{\mathcal H}=-\tilde{N}\left(\omega C^*C-{\mathcal E}\right).
\end{eqnarray}
The classical dynamics of these variables in the $\tilde{N}=1/\omega$ gauge is
$C=\sqrt{\frac{{\mathcal E}}{\omega}}\exp(i\eta)$. Moreover, consider on $\Gamma$ the functions
\begin{eqnarray}\label{22}
\begin{cases}
J_0:=\frac{1}{2}C^*C,\\
J_+:=\frac{1}{2}C^{*2},\\
J_-:=\frac{1}{2}C^2,
\end{cases}
\end{eqnarray}
which have a closed algebra
\begin{eqnarray}\label{23}
\begin{cases}
\{J_0,J_{\pm}\}=\mp iJ_{\pm},\\
\{J_+,J_-\}=2iJ_0.
\end{cases}
\end{eqnarray}
Since the phase space is two dimensional, there will be at most two independent
constraints. The Hamiltonian constraint implies
\begin{eqnarray}\label{24}
J_0=\frac{{\mathcal E}}{2\omega}.
\end{eqnarray}
Furthermore, we have
\begin{eqnarray}\label{25}
J^2:=J_0^2-\frac{1}{2}(J_+J_-+J_-J_+)=j(j-1),
\end{eqnarray}
where $j=\{1/4,3/4\}$ denote the Bargmann indexes for the simple harmonic oscillator.
Recall that an observable is a function on $\Gamma$ whose Poisson brackets with the first class constraints vanish when the first class constraints hold \cite{gauge}. Note that 
\begin{eqnarray}\label{26}
\{J^2,J_0\}=0,
\end{eqnarray}
which consequently implies that the Bargmann index is a gauge invariant observable: $J^2$ has strongly vanishing Poisson bracket with Hamiltonian and its value is
a constant of motion.

\subsection{Hidden symmetry and boundary conditions}
The boundary conditions for the evolution of subsystems of the Universe
are obtained from observations  out of the subsystem; they are related
to the rest of the Universe. On the other hand, in quantum cosmology, there is
no rest of Universe to pass their specification off to. ``The cosmological boundary condition must be one of the fundamental
laws of physics'' \cite{Hartle} or, as we investigate herein,  it can be related, at least in some specific, albeit restrictive, circumstances,  to the constraint algebra  of the cosmological model. In  this subsection, we will   obtain  boundary conditions using the hidden
dynamical symmetries of the model. 
To do this, we focus our attention on the Dirac observables
of the cosmological model. 

Let us start by introducing
the set of operators
$\hat{S}=\{C,C^\dagger,1\}$, which will have the commutator algebra
\begin{eqnarray}
[C,C^\dagger]=1,\hspace{.2cm}[C,1]=[C^\dagger,1]=0.
\end{eqnarray}
Hence, the set $\hat S$ and its commutator algebra are the quantum counterpart
of the set $S$.
The action of operators $\{C,C^\dagger\}$ on the states of the physical Hilbert
space are given by
\begin{eqnarray}\label{a}
\begin{cases}
C|n>=\sqrt{n}|n-1\rangle,\\
C^\dagger|n>=\sqrt{n+1}|n+1\rangle.
\end{cases}
\end{eqnarray}
 The Poisson bracket algebra of the classical $J$ 's can subsequently be promoted into a commutator algebra version by setting
\begin{eqnarray}\label{33}
\begin{cases}
J_0:=\frac{1}{4}(C^\dagger C+CC^\dagger),\\
J_+:=\frac{1}{2}C^{\dagger2},\\
J_-:=\frac{1}{2}C^2,
\end{cases}
\end{eqnarray}
so that the corresponding commutators are
\begin{eqnarray}\label{34}
[J_+,J_-]=-2J_0,\hspace{.2cm}[J_0,J_{\pm}]=\pm J_{\pm}.
\end{eqnarray}
Note that the above relations are recognized as the commutators of the Lie algebra of $su(1,1)$.
The positive discrete series representations of this Lie algebra are labeled by a positive real number $j>0$ (the Bargmann index).
The action of the above generators on a set of basis eigenvectors $|j,m\rangle$
are given by
\begin{eqnarray}\label{35}
\begin{cases}
J_{0}|j,m\rangle=(j+m)|j,m\rangle,\\
J_{+}|j,m\rangle=\sqrt{(2j+m)(m+1)}|j,m+1\rangle,\\
J_-|j,m\rangle=\sqrt{m(2j+m-1)}|j,m-1\rangle,
\end{cases}
\end{eqnarray}
where $m$ can be any non-negative integer. The corresponding Casimir operator can be calculated as
\begin{eqnarray}\label{36}
\begin{cases}
J^2:=J_0(J_0+1)-J_-J_+,\\
J^2|j,m\rangle=j(j-1)|j,m\rangle,
\end{cases}
\end{eqnarray}
with the following properties
\begin{eqnarray}\label{37}
[J^2,J_{\pm}]=0,\hspace{.2cm}[J^2,J_0]=0.
\end{eqnarray}
Thus, a representation of $su(1,1)$ is determined by the number $j$ and the eigenstates of $J^2$ and $J_0$,  constituting a 
basis for the irreducible representations of $su(1,1)$ and can be labelled by $|j,m\rangle$.
In addition, the Hamiltonian can be presented as
\begin{eqnarray}\label{38}
H=-{\mathcal E}+\omega(C^\dagger C+\frac{1}{2})=-{\mathcal E}+2\omega J_0,
\end{eqnarray}
which leads us to  point that the Casimir operator commutes with the Hamiltonian\begin{eqnarray}\label{39}
[J^2,H]=0.
\end{eqnarray}
As $J^2$ and $J_0$ commute with the Hamiltonian, they leave the physical Hilbert
space $V_H$ invariant and consequently we choose  $\{J_0,J^2,1\}$ as physical operators of the model
. Using definition (\ref{33}), the Casimir operator of
$su(1,1)$ reduces identically to $J^2=j(j-1)=-3/16$. Hence, the Bargmann index $j=\{\frac{1}{4},\frac{3}{4}\}$ is a gauge invariant
observable of the quantum cosmological model.
As a consequence,  from (\ref{13}), (\ref{35}) and (\ref{38}) we obtain
\begin{eqnarray}\label{40}
{\mathcal E}_{m,j}=2\omega(j+m).
\end{eqnarray}
Hence, the states of the Hilbert space, by means of the Hamiltonian constraint $V_{H=0}$, can be classified   in terms of the Bargmann index, allowing to
establish   two invariant subspaces:
\begin{eqnarray}\label{41}
\begin{cases}
{\mathcal E}_{\frac{3}{4},m}=\omega(\frac{3}{2}+2m);\hspace{.2cm}V_{H=0,j=\frac{3}{4}}=\{|\frac{3}{4},m\rangle\},\\
{\mathcal E}_{\frac{1}{4},m}=\omega(\frac{1}{2}+2m);\hspace{.2cm}V_{H=0,j=\frac{1}{4}}=\{|\frac{1}{4},m\rangle\},
\end{cases}
\end{eqnarray}
with $V_{H=0}=V_{H=0,j=\frac{1}{4}}\oplus V_{H=0,j=\frac{3}{4}}$.
Therefore, the gauge invariance of the Bargmann index  implies  a partition  of the Hilbert
space into  two disjointed invariant  subspaces,
 which are equivalent to the result of imposing boundary conditions (\ref{29}) and (\ref{29a}),  respectively.

\section{Conclusion and discussion}
We investigated how  the selection of boundary proposals in  quantum cosmology
can be related to the Dirac observables. In this paper, we have extracted
 Dirac observables of a closed Friedmann Universe, where the matter content is
constituted by non-interacting
radiation and dark matter (dust) perfect fluids. The reduced phase
space quantization
of this simple cosmological model was discussed. It  was shown that the hidden symmetry
of model, $su( 1,1)$, admits a Dirac observable related to  boundary proposals admissible for the model.

Notwithstanding the interest that the above paragraph may raise, the following
should be added:

\begin{enumerate}
\item 
 Our simple model is very specific, either in geometry or matter content choice. A wider analysis, with less restrictive cosmologies (but still bearing some symmetries) and/or other matter fields, should follow. The presence of fluid matter (as in (\ref{2})) was broadly used in, e.g., \cite{Ryan} so
that exact solutions of the (simplified) WDW equation could be obtained
(c.f. eq. (\ref{30})). Using instead a scalar field, e.g., would be more
generic and  more realistic from point of view of matter interaction
with the gravitational field in a high energy regime, where quantum effects
can be expected. Our proposal is that the presence of a hidden symmetry (as
herein denoted within the algebra of observables) is paramount to support the claim in the abstract; that from the algebra of constraints, (some) reasonable boundary conditions can be suitably extracted. We  suggest this could be verified within models where symmetries (like string dualities 
\cite{C,D,E,F}), acting directly, intertwining  geometrical elements and
matter field, are implicity present. We are addressing this 
in a forthcoming paper, considering scalar-tensor theories in a string setting
\cite{E,F,G}.  
\item
 The herein argumentation relies on the fact that the model has a singularity, as
mentioned at the end of section II. This implies a concrete to a domain of existence
for the wave function, $\psi$, and subsequently, requiring the Hamiltonian
to be unequivocally self-adjoint. This allows the boundary conditions pointed out in
IIIA and then the reasoning indicated in IIIC, from the algebra of constraints. Nevertheless, other boundary conditions can be put forward (e.g., \cite{HH}
or \cite{Vilenkin}), with well known results, which have been widely investigated
in the literature, including consistency and potential observational features\cite{A,B,HH,Vilenkin}.
It would be interesting to find some algebraic support for them, along the
lines discussed in section III, but the statements defining the boundary
conditions in \cite{HH,Vilenkin} are more of a ``topological'' nature for
the mini-superspaces involved and hence it is not obvious if this can be
achieved.  Moreover, singularities of a differential nature can be present
in a given cosmology (e.g., late time \cite{gamma1} or pre-Big-Bang \cite{C,D})
and a discussion involving them, conditions on $\psi$ and admissible boundary conditions
by means of the algebra of constraints is worthy, namely if within the context
of hidden symmetries.
\item Finally, it can be of interest to point  the (partial) similarities of
Eqs. (\ref{20})-(\ref{22}) or (\ref{33})-(\ref{38}) (regarding $J_0$, the
definition of $C$, $C^\dagger$
with respect to same elements present in supersymmetric quantum mechanics
\cite{delta1,delta2,delta3,E,F}. In fact, $J_0$ in (\ref{33}) suggests a
anti-commutation relation, whereas from (\ref{20}) we can infer (part of)
a $N=2$ super-charge structure in the same manner as in \cite{delta1,delta2,delta3,E,F,Sp}.
These assertions need  to be carefully explored, but we think that a
possible relation between the degeneracy from (\ref{40}), together with the
relation between boundary conditions, self-adjoint Hamiltonian, classical
singularities, regarding  the integrability of a WDW equation, is worthy
to explore further by means of broader cosmologies. 
 \end{enumerate}

\end{document}